\documentclass[unnumsec,contemporary,large]{oup-authoring-template}%

\theoremstyle{thmstyleone}

\theoremstyle{thmstyletwo}

\theoremstyle{thmstylethree}

\usepackage{booktabs}

\usepackage{natbib}
\setcitestyle{sort&compress}
\usepackage{pifont}
\usepackage{xcolor}

\newcommand{\gcheck}{\textcolor{green!60!black}{\ding{51}}}
\newcommand{\rx}{\textcolor{red!70!black}{\ding{55}}}

\newif\ifshowsupplemental
\showsupplementaltrue  

\begin{document}

\journaltitle{Bioinformatics}
\DOI{DOI HERE}
\copyrightyear{2025}
\pubyear{2025}
\access{Advance Access Publication Date: Day Month Year}
\appnotes{Application Note}

\firstpage{1}


\title[AmpliconHunter2]{AmpliconHunter2: a SIMD-Accelerated $\textit{In-Silico}$ PCR Engine}

\author[1,$\ast$]{Rye Howard-Stone\ORCID{0009-0005-7518-1867}}
\author[1]{Ion I. M\u{a}ndoiu\ORCID{0000-0002-4818-0237}}

\authormark{Howard-Stone and M\u{a}ndoiu}

\address[1]{\orgdiv{School of Computing}, \orgname{University of Connecticut}, \orgaddress{\street{371 Fairfield Way}, \postcode{06269}, \state{CT}, \country{USA}}}

\corresp[$\ast$]{Corresponding author. \href{mailto:rye.howard-stone@uconn.edu}{rye.howard-stone@uconn.edu}}

\received{Date}{0}{Year}
\revised{Date}{0}{Year}
\accepted{Date}{0}{Year}


\abstract{
\textbf{Summary:} We present AmpliconHunter2 (AHv2), a highly scalable \emph{in silico} PCR engine written in C that can handle degenerate primers and uses a highly accurate melting temperature model. AHv2 implements a bit-mask IUPAC matcher with AVX2 SIMD acceleration, supports user-specified mismatches and 3’ clamp constraints, calls amplicons in all four primer pair orientations (FR/RF/FF/RR), and optionally trims primers and extracts fixed-length flanking barcodes into FASTA headers. The pipeline packs FASTA into 2-bit batches, streams them in 16 MB chunks, writes amplicons to per-thread temp files and concatenates outputs, minimizing peak RSS during amplicon finding. We also summarize updates to the Python reference (AHv1.1). \\
\textbf{Availability and Implementation:} AmpliconHunter2 is available as a freely available webserver at: \url{https://ah2.uconn.engr.edu}. Source code is available at: \url{https://github.com/rhowardstone/AmpliconHunter2} under an MIT license. AHv2 was implemented in C; AHv1.1 using Python 3 with Hyperscan. \\
\textbf{Contact:} \href{mailto:rye.howard-stone@uconn.edu}{rye.howard-stone@uconn.edu} \\
\textbf{Supplementary information:} Supplementary data are available at \textit{Bioinformatics} online.
}


\maketitle

\section{Introduction}

Polymerase chain reaction (PCR) amplicon sequencing remains a cornerstone of microbiome profiling because it is cost-effective, scalable and can generate taxonomic profiles from complex microbial communities. Most microbial surveys amplify portions of the 16S rRNA gene because this gene contains nine hypervariable regions (V1--V9) flanked by conserved segments that allow the same primers to amplify diverse bacterial taxa. Despite its ubiquity, amplicon sequencing can be affected by amplification bias: variability in primer binding sites across taxa means that some organisms may be under-represented in sequence data. Indeed, recent studies emphasize that commonly used universal primer sets often fail to capture the full microbial diversity in a sample and that primer design must account for inter-genomic variation \citep{sunthornthummas2025}.

As \emph{in vitro} experimentation is expensive, practitioners frequently use \emph{in silico} tools to predict genome amplification patterns and assess off-target amplification before wet-lab experiments. Existing \emph{in silico} tools, however, struggle to process currently available million-genome datasets or lack features such as the accurate modeling of melting temperature with mismatches required by commonly used sets of degenerate primers. To address these shortcomings, we recently released AmpliconHunter v1 (AHv1), a scalable \emph{in silico} PCR package implemented in Python \citep{ampliconhunter}. AHv1 relies on the Hyperscan regex engine \cite{Wang2019} for performing approximate matching with mismatches and performs nearest-neighbor melting temperature calculations using the BioPython's Tm\_NN function.

In this paper we further improve the scalability of \emph{in silico} PCR by introducing AmpliconHunter v2 (AHv2). AHv2 is written in C and implements a bit-mask IUPAC matcher with AVX2 SIMD acceleration. For this tool, we also implemented a C version of the nearest-neighbor melting temperature model that is identical to BioPython's Tm\_NN function. In the rest of the paper we describe AHv2 and provide benchmarking results comparing it with an updated version of AHv1 (AHv1.1) and several interim versions (Table~\ref{tab:features}).



\section{Implementations}
\textbf{AHv1.1} is a functional update to AmpliconHunter \citep{ampliconhunter}, that permits FASTQ input and output, optionally trims primer sequences, extracts fixed-length flanking barcodes and can automatically filter off-target amplicons. When published, AmpliconHunter was compared against prior efforts such as PrimerEvalPy \citep{VazquezGonzalez2024} and Ribdif2 \citep{Murphy2023}, which offer utilities for \emph{in silico} PCR, but are not designed for million-genome scale. AHv1.1 retains nearly identical runtime characteristics as version 1, which was found to be 100 times faster than PrimerEvalPy and 10 times faster than Ribdif2, while supporting significantly more features.

\textbf{AHv2.$\alpha$} was rewritten in C to use 2-bit compressed data for performance. Input FASTA files are compressed into 2-bit batches that store headers, lengths and packed bases. These batches are memory-mapped with \texttt{MAP\_PRIVATE}; AHv2.$\alpha$ advises the kernel that pages will be accessed sequentially and dropped once consumed using \texttt{posix\_madvise} with \texttt{POSIX\_MADV\_SEQUENTIAL} and \texttt{POSIX\_MADV\_DONTNEED}
The engine enforces a user-defined 3-prime clamp, streams amplicons to per-thread buffers and periodically merges them to minimize peak resident set size (RSS). AHv2.$\alpha$ outputs FASTA only and omits melting temperature ($T_m$), HMM calculation, decoy analysis, and taxonomy modules for simplicity. 

\textbf{AHv2.$\beta$} builds on AHv2.$\alpha$ by incorporating AVX2 parallelization, improved thread scheduling and dynamic buffer allocation. Primers are compiled into per-base IUPAC masks and matched using AVX2 256-bit registers; each register holds 32 bytes, allowing mismatches to be counted in parallel. AHv2.$\beta$ achieves lower runtime but consumes untenable RAM.

\textbf{AHv2.$\gamma$} further optimizes the AVX2 matcher and I/O pipeline by reading data in 16~MB chunks. It hoists primer masks into contiguous vectors, employs unrolled loops to reduce branch mispredictions, and aggressively reclaims memory after each batch.  AHv2.$\gamma$ also uses static scheduling to mitigate load imbalance. These changes yield the fastest runtimes and the best parallel efficiency but still incur a moderate memory footprint due to decoded cache blocks.

\textbf{AHv2} reimplements BioPython's entire Tm\_NN function in C, permitting exact replication of the calculations in AHv1.1, for any parameter combination and all salt correction methods. Results are rounded to two decimal places but are otherwise identical to the BioPython annotations. AHv2 omits HMMs, decoys, taxonomy, and FASTQ features by design for minimal dependencies and maximal throughput.

\section{Results}

We compared AHv1.1--AHv2 using six tests. All comparisons used 204.8K genomes from the AllTheBacteria genome collection, a set of 2.4M publicly available bacterial assemblies that we previously used for large-scale amplicon evaluation \citep{Hunt2024AllTheBacteria}. Instructions for downloading the genomes we tested are shown as part of our supplementary repository. Each test includes multiple replicates.
The benchmarking script used the same set of primer pairs (V1V9) and parameters (two mismatches, clamp size of 3) across versions. Tests 1--5 were all conducted on the 12.8K genome subset. Filtering based on $T_m$ is disabled in both implementations that support it: AHv1.1 and AHv2; however, both implementations automatically compute and annotate amplicons with $T_m$ using the same nearest-neighbor model.

\begin{figure*}[t]
  \centering
  \includegraphics[width=0.9\textwidth]{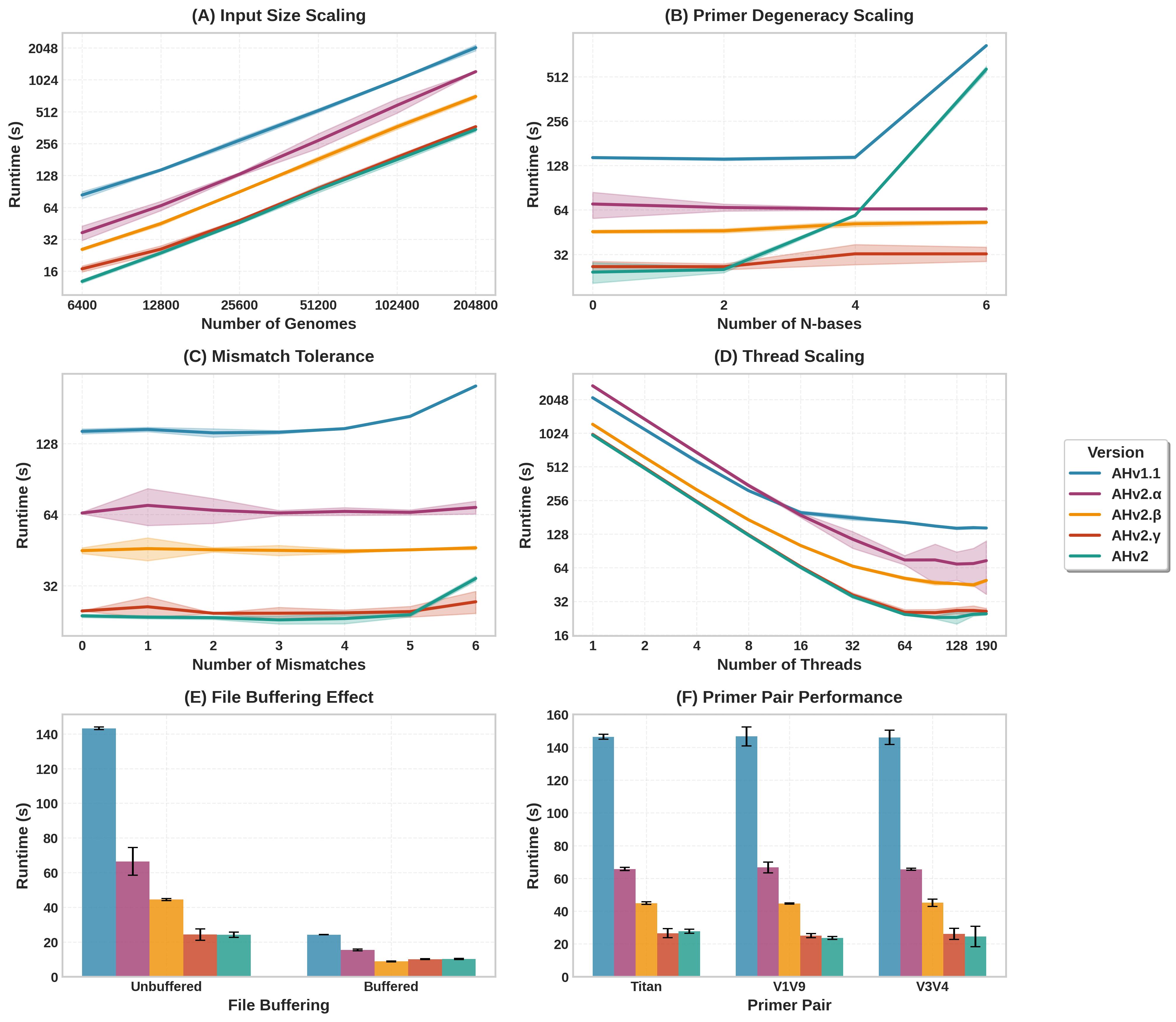}
  \caption{Runtimes for AHv1.1–AHv2 across six experiments. Lines are means; ribbons show 95\% CI.  (A) Input-size scaling across 6.4K–204.8K genomes. (B) Effect of primer degeneracy (number of appended N bases). (C) Effect of allowed mismatches (substitutions). (D) Thread scaling from 1–190 threads. (E) Cold vs warm cache performance. (F) Primer pair performance (Titan, V1V9, V3V4). }

  \label{fig:runtime}
\end{figure*}

\subsection{Input size scaling}

Figure~\ref{fig:runtime}A shows the runtime as the number of genomes increases (6.4K--204.8K). AHv2 consistently performs the fastest, followed by AHv2.$\gamma$, AHv2.$\beta$, AHv2.$\alpha$ and AHv1.1. AHv2 completes 204.8K genomes in 347.9~seconds (95\% CI: [332.0, 363.8]), whereas AHv1.1 requires 2056.5~seconds (95\% CI: [1925.5, 2187.6]).

Memory usage (Figure~\ref{fig:S1}A) tells a different story: AHv2.$\beta$ uses significantly more RAM (45.8~GB at 204.8K genomes). AHv2 and AHv2.$\gamma$ maintain moderate memory usage around 3.9~GB. AHv2.$\alpha$ uses $\sim$0.73~GB, while AHv1.1 maintains the lowest RSS regardless of input size (0.48~GB for 204.8K genomes).

Scaling efficiency (Figure~\ref{fig:S1}B) is computed as runtime for a baseline input divided by runtime for larger inputs normalized by input size. Perfect scaling would remain at 100\%. AHv2.$\gamma$ exceeds 140\% efficiency at intermediate sizes. AHv1.1's efficiency increases to $\sim$130\% at the largest input size, suggesting its better-than-linear scaling for the Python implementation is due to amortized overhead. AHv2 and AHv2.$\beta$ peak just below 120\%, while AHv2.$\alpha$ descends slightly beneath 100\% at large inputs.

System-level metrics also reveal differences. All AHv2 variants read only a fraction of the input volume compared with AHv1.1 because they use compressed 2-bit batches, whereas AHv1.1 reads entire FASTA files (approximately 3.93$\times$ reduction in filesize). Context switching (Figure~\ref{fig:S1}C) increases with the number of genomes, with AHv2.$\alpha$ showing the highest count at large genome sizes. AHv1.1's system-time fraction (Figure~\ref{fig:S1}D) decreases as the input grows ($\sim$7\% at 6.4K genomes), while all C implementations maintain consistently low system overhead (1--4\%). 

\begin{table}[t]
\centering
\small
\begin{tabular}{@{}lccccc@{}}
\toprule
\textbf{Implementation Detail} & \textbf{v1.1} & \textbf{v2.$\alpha$} & \textbf{v2.$\beta$} & \textbf{v2.$\gamma$} & \textbf{v2} \\
\midrule
Python implementation            & \gcheck & \rx    & \rx    & \rx    & \rx    \\
Hyperscan regex                  & \gcheck & \rx    & \rx    & \rx    & \rx    \\
SIMD acceleration              & \gcheck*     & \rx& \gcheck$\dagger$& \gcheck$\dagger$& \gcheck$\dagger$\\
C implementation            & \rx     & \gcheck& \gcheck& \gcheck& \gcheck\\
2-bit encoded pipeline           & \rx     & \gcheck& \gcheck& \gcheck& \gcheck\\
Memory-mapped I/O          & \rx     & \gcheck& \gcheck& \rx& \rx\\
Buffered I/O          & \gcheck     & \rx& \rx& \gcheck& \gcheck\\
Bit-mask IUPAC           & \rx     & \gcheck& \gcheck& \gcheck& \gcheck\\
Aggressive buffer reclamation    & \rx     & \rx    & \rx    & \gcheck& \gcheck\\
Melting Temperature calculation    & \gcheck     & \rx    & \rx   & \rx    & \gcheck\\
\bottomrule \\
\end{tabular}
\caption{Binary feature and implementation matrix across AmpliconHunter versions (green check = supported, red X = not supported). Rows with identical values across all versions were removed.  *AHv1.1 utilizes Hyperscan which implicitly uses SIMD acceleration. $\dagger$ AmpliconHunter2 explicitly uses AVX2 instructions.}
\label{tab:features}
\end{table}

\subsection{Primer degeneracy and mismatch tolerance}

Degenerate primer bases (``N'' positions) increase the size of the search space. In Figure~\ref{fig:runtime}B, two N bases are added at a time (one per primer) and used to extract amplicons from the 12.8K genome subset. AHv1.1's runtime increases exponentially when primers contain six degenerate bases (reaching over 830~seconds), whereas the earlier C implementations maintain nearly constant runtime (AHv2.$\gamma$ remains around 30~seconds). This highlights the advantage of bit-mask matching: AVX2 vectorizes IUPAC comparisons and avoids the combinatorial explosion inherent in regex matching. However, AHv2 also exhibits an exponential increase in response to increased degeneracy in input primers, as $T_m$ needs to be calculated for more primer variants. Memory usage (Figure~\ref{fig:S2}A) is largely unaffected by degeneracy for all implementations.

Mismatch tolerance has similar effects on the search space (Figure~\ref{fig:runtime}C). Allowing more mismatches slows all versions, but the C versions degrade more gracefully because mismatches can be counted efficiently using bitwise operations, whereas Hyperscan must evaluate many patterns. At six mismatches AHv1.1's runtime increases to approximately 224~seconds, while AHv2 remains at 34.4~seconds. Memory footprints (Figure~\ref{fig:S2}B) are largely flat, with a slight increase for AHv1.1 and AHv2.$\alpha$ at six allowed mismatches.

\subsection{Thread scaling and parallel efficiency}

Figure~\ref{fig:runtime}D reports runtime as the number of threads increases (1--190). AHv2 achieves near-linear speed-ups up to 64 threads before saturating, completing the largest dataset in $\sim$24.8~seconds at 190 threads. AHv2.$\gamma$ scales similarly well, reaching about 26~seconds at 190 threads. AHv2.$\beta$ scales well to 32 threads but plateaus thereafter just below 50~seconds. AHv2.$\alpha$ stabilizes around 74~seconds. AHv1.1 shows the poorest parallel scaling; beyond 16 threads additional cores provide diminishing returns, leveling off at $\sim$144~seconds.

Parallel efficiency, the ratio of ideal to observed runtime scaling relative to a baseline input, is shown in Figure~\ref{fig:S2}C. AHv2 and AHv2.$\gamma$ maintain over 90\% efficiency up to 16 threads and remain above 80\% at 32 threads, but drop afterwards. AHv1.1 drops below 90\% with as few as 8 threads, plummeting to less than 40\% efficiency with 32 threads.
CPU utilization (Figure~\ref{fig:S2}D) further illustrates these trends: AHv2, AHv2.$\gamma$ and AHv2.$\beta$ saturate cores, reaching over 4,000\% utilization (about 40 cores $\times$ 100\%), while AHv1.1 remains under 1,500\% likely because Python's Global Interpreter Lock serializes some operations.

\subsection{Cache performance}

Figure~\ref{fig:runtime}E and Figure~\ref{fig:S2}E evaluate cold versus warm cache conditions. When the OS page cache is cold (first run), AHv1.1 spends most of its time in disk I/O, taking $\sim$140~seconds. AHv2.$\alpha$ completes in $\sim$70~seconds, AHv2.$\beta$ in $\sim$45~seconds, AHv2.$\gamma$ in $\sim$25~seconds and AHv2 in 22~seconds. On repeated runs (warm caches), all implementations improve, but AHv2.$\beta$ improves the most, becoming competitive with AHv2.$\gamma$ and AHv2. The cache speed-up factor (Figure~\ref{fig:S2}E) quantifies this improvement: AHv1.1 gains just shy of a 6$\times$ speed-up from caching, AHv2.$\alpha$ roughly 4.3$\times$, AHv2.$\beta$ about 5$\times$, AHv2.$\gamma$ approximately 2.4$\times$ and AHv2 about 2.3$\times$.

\subsection{Primer-pair performance}

We evaluated three commonly used primer pairs (Titan, V1V9 and V3V4) across all implementations (Figure~\ref{fig:runtime}F). For the Titan primer pair, AHv2 completes in 27.7~seconds, AHv2.$\gamma$ in 26.5~seconds, AHv2.$\beta$ in 44.9~seconds, AHv2.$\alpha$ in 65.7~seconds, and AHv1.1 in 146.4~seconds. Similar patterns hold for V1V9 and V3V4 primers with minimal variation between them. Memory usage was nearly identical between primer pairs for all methods. These results suggest that the level of degeneracy of primers and amplicon lengths (within practical ranges) have little effect on runtime beyond constant factors; performance differences are dominated by architecture and caching strategies.


\section{Discussion and future directions}
Overall, our results highlight a trade-off between speed and memory. AHv2 is the fastest implementation, with a speed-up factor of 5.91$\times$ over AHv1.1 at 204.8K genomes (Figure~\ref{fig:S2}F). Its optimizations and parallelizations make it the most suitable for large-scale projects where throughput is paramount. AHv2.$\gamma$ offers nearly equivalent speed (5.57$\times$ speed-up) but does not include melting temperature calculation. AHv2.$\beta$ provides a 2.89$\times$ speed-up but requires substantial memory (45.8~GB). AHv2.$\alpha$ strikes a balance, delivering a mere 1.68$\times$ speed-up over AHv1.1 but maintaining a tiny memory footprint (0.73~GB). AHv1.1 remains useful for its advanced features (HMMs, taxonomic summarization, and FASTQ support) but is not competitive at scale.

The move from a Python + regex engine to a C + AVX2 bit-mask matcher yields dramatic performance gains. Vectorized matching allows 32 bases to be compared in a single instruction, and streaming 2-bit batches reduces I/O and memory traffic. The results show that naive regex matching struggles with degenerate primers and mismatch tolerance, whereas bit-mask approaches scale gracefully.

However, the AHv2 C implementation does not include all original functionality. It currently supports FASTA input only, omitting HMM scoring, decoy sequences, and taxonomic summaries; these features remain available in AHv1.1. AHv2 scales well to dozens of cores, but scaling saturates beyond 64 threads due to I/O bottlenecks and memory contention; further improvements could involve explicit NUMA-aware scheduling and prefetching (non-uniform memory access).

Future releases should aim to close the functionality gap by incorporating modules for HMM scoring and decoys while preserving speed. Support for FASTQ input with quality propagation would make the C implementation usable for processing amplicon sequencing data. Separating sequence headers as part of 2-bit compression may additionally shave some small amount of time off execution. Exploring AVX-512 and ARM NEON intrinsics could also provide speed-ups on newer hardware. In addition, GPU-accelerated matching 
may offer further performance gains.

\section{Conclusion}

AmpliconHunter2 demonstrates that careful engineering, including vectorized IUPAC matching, 2-bit encoding, and careful memory management can significantly speed up demanding high-performance computing workflows. To increase the usability of AmpliconHunter2, we have released a corresponding webserver, as we did for AHv1. Users of the webserver are empowered to assess the quality of their primers against large microbial genomic collections (including subsets of RefSeq \citep{OLeary2016RefSeq} genomes and complete versions of the PATRIC \cite{Gillespie2011PATRIC}, GTDB \citep{Parks2021GTDB}, and AllTheBacteria databases \citep{Hunt2024AllTheBacteria}), without access to command-line tools or high-performance computing environments. Our results show that the AHv2 webserver completes the analysis for V1V9 primers on the $\sim$2.4M genomes from the AllTheBacteria project in 38.73 minutes, compared to 419.45 minutes for the AmpliconHunter webserver ($\sim$10.8x speedup). For reproducibility, we provide all intermediary implementations and benchmarking scripts on our supplementary GitHub page (\url{https://github.com/rhowardstone/AmpliconHunter2_benchmark}). Together, these implementations provide a flexible toolkit for microbiome researchers and primer designers operating at terabyte-scale.

\bibliography{references}

\ifshowsupplemental
\clearpage
\onecolumn
\setcounter{figure}{0}
\setcounter{table}{0}
\renewcommand{\thefigure}{S\arabic{figure}}
\renewcommand{\thetable}{S\arabic{table}}
\section*{Supplementary Information}

\subsection*{Benchmarking script and reproducibility}

The benchmarking script (\texttt{run\_benchmarks.sh}) iterates over the six tests shown in \ref{fig:runtime}. For each test, it runs all four versions of AmpliconHunter using comparable parameters, captures wall-clock time with \texttt{/usr/bin/time -v} and collects system metrics via \texttt{/proc/self/status} and \texttt{perf stat}. The script writes a JSON summary (\texttt{benchmark\_results.json}) that includes runtimes, peak RSS, page-fault counts, context switches, and CPU utilization for each replicate. Users can reproduce the plots by running \texttt{generate\_publication\_figures.py} on the JSON file. The repository also contains example primer files and a manifest of compressed genome batches.

\subsection*{Hardware and software environment}

Benchmarks were performed on an Ubuntu 22.04 server with dual Intel Xeon Gold 6248R CPUs (2\(\times\)24 cores, 3.0~GHz) and 512~GB of DDR4 memory. The C implementations were compiled with GCC 13.3 using \texttt{-O3}, \texttt{-march=native} and \texttt{-mavx2}. AHv1.1 used Python 3.11 and Hyperscan 0.7.8.

\subsection*{Data availability}

All code, compressed genome batches, benchmarking scripts and raw results are available from the accompanying GitHub repository (\url{https://github.com/rhowardstone/AmpliconHunter2_benchmark}). Figures were generated using the provided scripts and can be reproduced on any modern Linux server. The authors welcome contributions and feature requests.

\subsection{Webserver}

For ease of use, we make a webserver available serving AmpliconHunter2 on the same databases used for the original AmpliconHunter web interface, with a slightly improved design. Past jobs are now much easier to find: searchable, and filterable by database as well as status. Plots are largely the same, with the removal of the HMM and decoy plots, but we have added a taxonomy breakdown that is searchable as a table, and navigable as a tree. We have kept our amplitype pattern plots, but opted for a static png for easy transferability. We have included several interactive plots, including distributions for amplicon length, GC content, and melting temperature, along with the primer orientation breakdown. Please visit \url{https://ah2.engr.uconn.edu/} to view and submit AmpliconHunter2 jobs.

\subsection{Primer matching and clamp logic (AHv2)}
Primers are converted to per-base IUPAC bit masks; reverse-complement masks are precomputed. Each sequence is converted once to a mask array. We count mismatches via AVX2 bitwise AND operations, with exact 3’ clamp enforced.

\subsection{Amplicon calling and orientation (AHv2)}
We sort candidate sites and pair opposite-sense hits within user bounds (\texttt{--min-length}, \texttt{--max-length}). We stop when a same-sense site appears (prevents invalid overlaps). Orientation codes are the same as version 1: FR, RF, FF, RR. By default, we emit FR+RF; \texttt{--include-offtarget} additionally emits FF/RR. RF amplicons are reverse-complemented so sequences are in forward orientation. 

\subsection{Primer trimming and barcode extraction (AHv2)}
With \texttt{--trim-primers}, we remove matched primer sequences from the emitted amplicon. Barcodes are fixed-length flanks upstream of the forward primer (\texttt{--fb-len}) and downstream of the reverse primer (\texttt{--rb-len}) for FR; the RF case extracts on the opposite sides and reverse-complements both barcodes.

\subsection{Headers and outputs (AHv2)}
Output is FASTA. Headers encode source file, genomic coordinates, orientation, matched primer snippets, and optional barcodes, e.g.:
\begin{verbatim}
>seqid.source=GCF_XXXX.fa.coordinates=12345-13567.Tm=60.42.orientation=FR
  .fprimer=... .rprimer=... .fb=ACGT ... .rb=TGCA
\end{verbatim}
(\texttt{.fb}/\texttt{.rb} only when requested.)

\newpage
\section*{Supplementary Figures}

\begin{figure*}[h]
    \centering
    \includegraphics[width=0.85\textwidth]{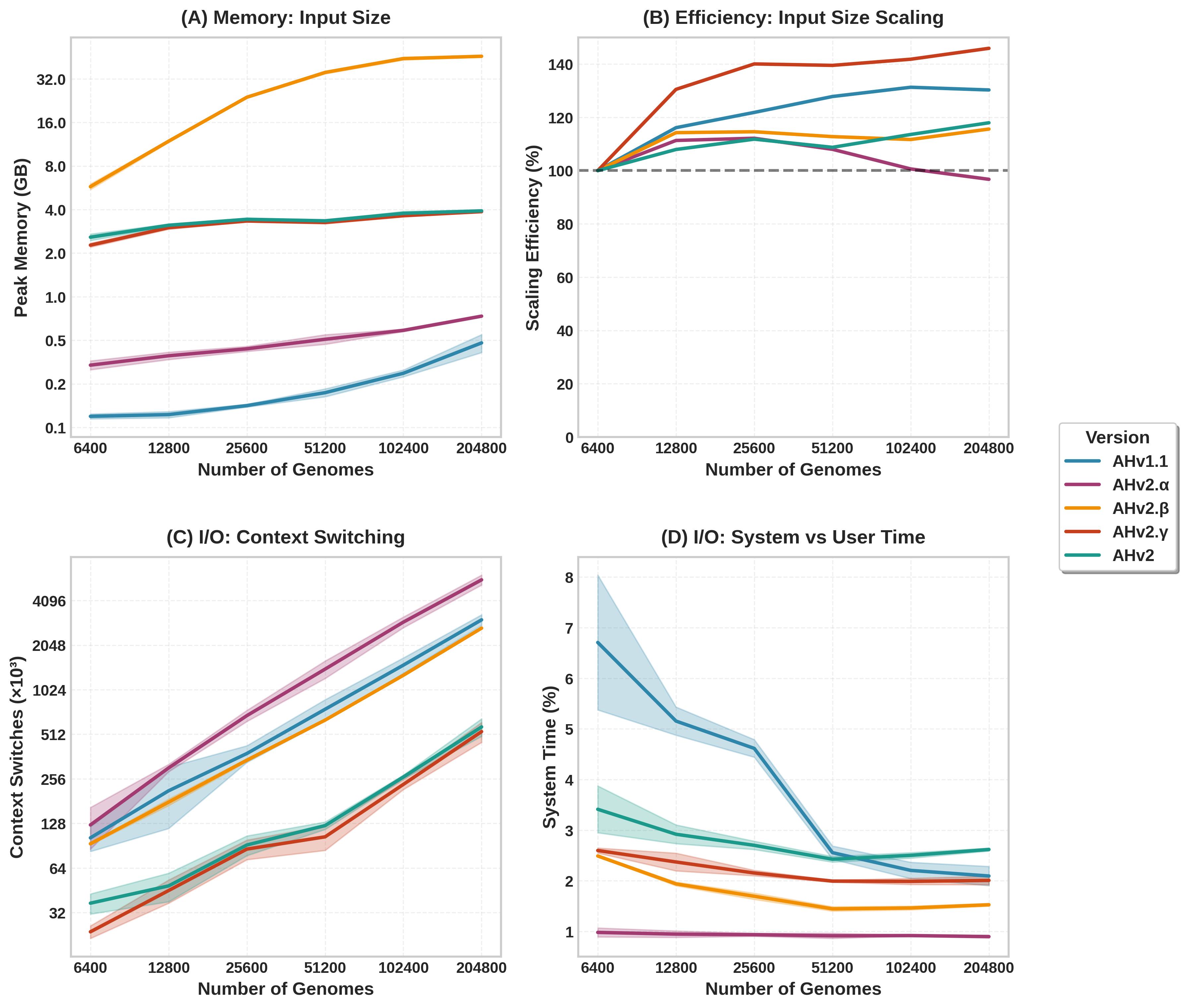}
    \caption{Memory and efficiency analysis for input size scaling. Panel (A) shows peak resident memory (GB) versus input size, (B) input size scaling
  efficiency, (C) context switching overhead, and (D) ratio of system time to user time. Means with replicate variability shown (95\% CI).}
    \label{fig:S1}
  \end{figure*}

  \begin{figure*}[h]
    \centering
    \includegraphics[width=\textwidth]{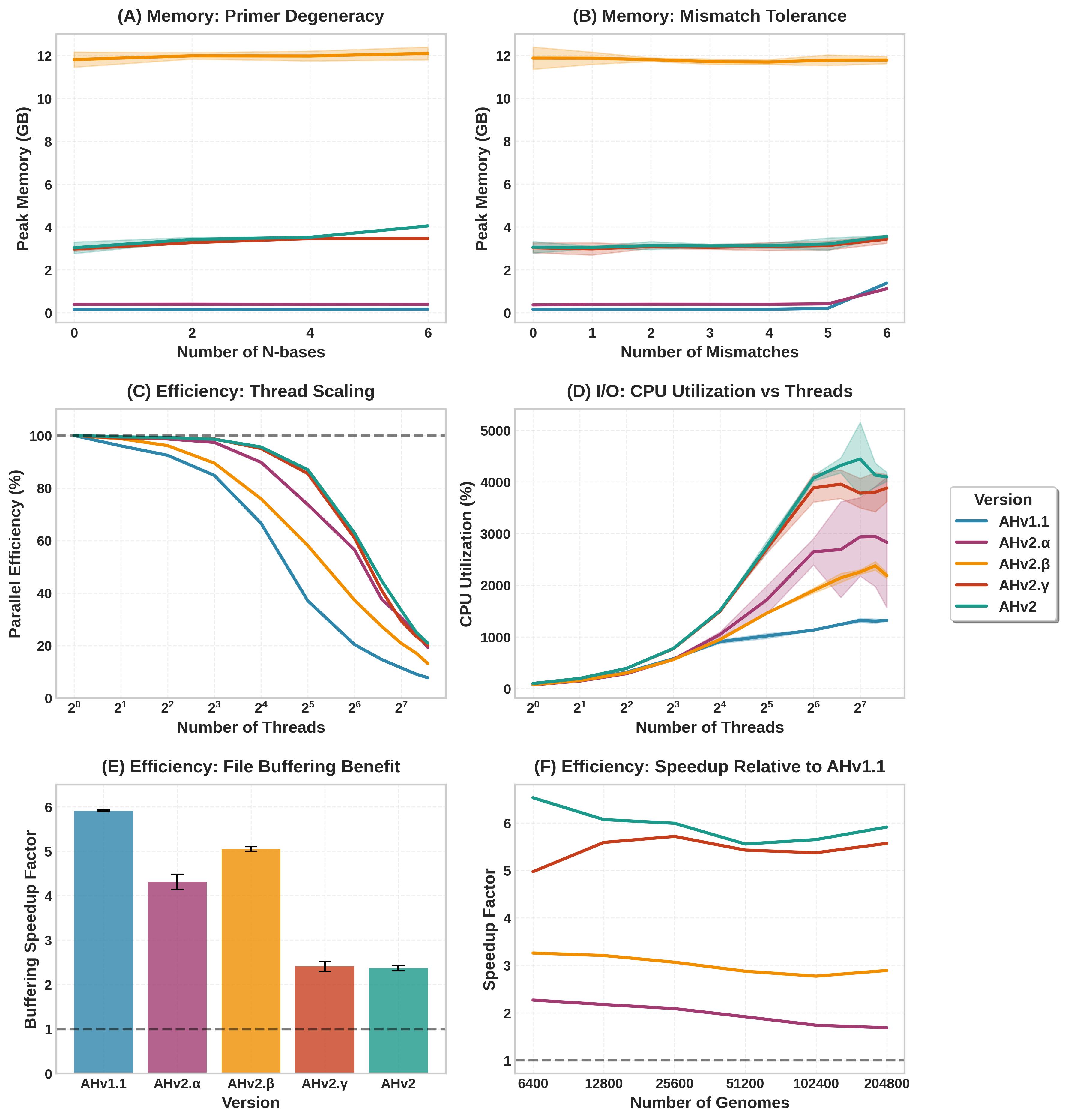}
    \caption{Extended performance analysis. Panel (A) shows peak memory for primer degeneracy, (B) peak memory for mismatch tolerance,
  (C) parallel efficiency versus thread count, (D) CPU utilization versus thread count, (E) cache performance benefit, and (F) speed-up
  relative to AHv1.1. Means with replicate variability shown (95\% CI).}
    \label{fig:S2}
  \end{figure*}

\fi

\end{document}